\documentstyle[aps,prd,preprint,epsf]{revtex}

\newcommand{\be}{\begin{equation}}
\newcommand{\ee}{\end{equation}}
\newcommand{\bea}{\begin{eqnarray}}
\newcommand{\beas}{\begin{eqnarray*}}
\newcommand{\eea}{\end{eqnarray}}
\newcommand{\eeas}{\end{eqnarray*}} 
\newcommand{\ba}{\begin{array}}
\newcommand{\ea}{\end{array}}
\newcommand{\bi}{\begin{itemize}}
\newcommand{\ei}{\end{itemize}}
\newcommand{\ben}{\begin{enumerate}}
\newcommand{\een}{\end{enumerate}}

\def\IC{\relax\hbox{$\inbar\kern-.3em{\rm C}$}}
\def\IQ{\relax\hbox{$\inbar\kern-.3em{\rm Q}$}}
\def\IR{\relax{\rm I\kern-.18em R}}

\def\IN{\relax{\rm I\kern-.18em N}}

 \font\cmss=cmss10 \font\cmsss=cmss10 at 7pt
\def\IZ{\relax\ifmmode\mathchoice
{\hbox{\cmss Z\kern-.4em Z}}{\hbox{\cmss Z\kern-.4em Z}}
 {\lower.9pt\hbox{\cmsss Z\kern-.4em Z}}
 {\lower1.2pt\hbox{\cmsss Z\kern-.4em Z}}\else{\cmss Z\kern-.4em Z}\fi}

\begin{document}

%\draft
%@twocolumnfalse\endcsname

\preprint{\vbox{
\hbox{UMD-PP-99-107}
}}

%\vskip1pc

\title{Effect of Extra Dimensions on Gauge Coupling Unification}

\author{A. P\'erez--Lorenzana$^{a,b}$ and R. N. Mohapatra$^a$}
\address{
$^a$ Department of Physics, University of Maryland, College Park, Maryland 
20742, USA\\
$^b$ Departamento de F\'\i sica, 
Centro de Investigaci\'on y de Estudios Avanzados del I.P.N.\\
Apdo. Post. 14-740, 07000, M\'exico, D.F., M\'exico.}

\maketitle

\begin{abstract}

The effects of extra dimensions on gauge coupling unification is studied.
We start with a comparison between power law running
of the gauge couplings in models with extra
dimensions and logarithmic running that happens in many realistic
cases. We then discuss the effect of extra dimensions on various classes
of unification models. We identify products of evolution coefficients that
dictate the profile of unification in different models. We use them to
study under what conditions unification of couplings can occur in both
one and two step unification models. We find that Kaluza-Klein modes
can help generate interesting intermediate scale models with gauge
coupling unification such as the minimal left-right models with the seesaw
mechanism with a $M_{W_R}\sim 10^{13}$ GeV intermediate scale, useful in 
understanding neutrino oscillations. We also obtain several examples
where the presence of noncanonical normalization of couplings enables us
to obtain unification scales around $10^{11}$ GeV. This fits very well
into a class of models proposed recently where the string scale is
advocated to be at this value from physical arguments.\\[1ex]

\end{abstract}

%\pacs{PACS:}
%\vskip2em]

%%%%%%%%%%%%%%%%%%%%%%%%%%%%%%
\section{Introduction}
%%%%%%%%%%%%%%%%%%%%%%%%%%%%%%
The idea that there may be more than three space dimensions is as old as
Kaluza and Klein's work dating back to the early part of this century.
The advent of superstring theories generated new interest in extra
dimensions since consistent superstring theories exist only in 10 or 26
dimensions. It was conventional to assume that the extra dimensions
are compactified to manifolds of small radii so that they remain hidden
to physics considerations. The value of the small radii was assumed to be 
of order $M^{-1}_{P\ell}$ and thus invisible.

Recently the possibility that the extra hidden dimensions may have
radii considerably larger (of order TeV$^{-1}$ or even (milli-eV)$^{-1}$)
has been the subject of intense scrutiny
~\cite{anto,dvali,pheno,dine,ddg1,ddg2,ross,kaku,carone,quiros,frampton}
and has generated considerable amount of excitement in phenomenological
circles. This is related to theoretical developments in string theories
which have made such speculations plausible. It is assumed that the extra
dimensions of string theories are compactified on orbifolds (or other
compact manifolds) of radii $R$. The sizes of these radii depend on
the details of the theory. In the currently popular pictures where it is 
assumed that there exist D-branes embedded in the high dimensional space, 
if one assumes a scenario where only gravity is in the bulk and all matter
fields are in the brane, $R$ can be as large as a milli meter\cite{dvali},
which is the threshold below which the Newtonian gravitational law has not
been experimentally verified. In other cases, $R$ has to be smaller than 
a few TeV$^{-1}$. This latter case may not only have direct 
experimental tests in colliders\cite{coll}, but also may have interesting
implications for physics beyond the standard model because its presence
crucially effects the nature of grand unification of forces and matter.
These ideas may also have other theoretical implications such
as a  new route to solve the hierarchy problem~\cite{dvali} if it is
assumed that the string scale is in the TeV range\cite{lykk}. Such low
scales also
raise possibilities for new effects in
astrophysical settings, which then lead to new constraints on 
them~\cite{pheno}. Clearly a rich new avenue of
particle physics has been opened up by these considerations.

In this article, we explore the effects of extra dimensions on the
unification of gauge couplings. Dienes, Dudas and Gherghetta\cite{ddg1}
began this kind of analysis in a series of papers for the minimal
supersymmetric standard model. They used power law unification, noted
originally by Taylor and Veneziano\cite{taylor} to argue
that indeed MSSM leads to unification even in the presence of extra hidden
dimensions with an arbitrary scale between a TeV$^{-1}$ and the inverse of
the GUT scale. Subsequent papers have addressed various issues related to
the question of unification~\cite{ross,kaku,carone,quiros,frampton}. 
For example, it has been noted that in the ``minimal'' versions of the
model considered in \cite{ddg1,ddg2}, the true unification of couplings
predicts a larger value for $\alpha_{s}(M_Z)$ compared to experimental
observations. One way to cure this problem is to include new contributions 
to beta functions\cite{ross,kaku,carone,quiros,frampton} by postulating
additional fields at the weak scale.

It is the goal of this paper to continue this investigation further. We
start by discussing briefly the power law variation of couplings
and comparing it with the logarithmic one that is closer to
reality in most unification models\cite{ross}. 

We then introduce a new set of variables
constructed out of the beta function coefficients and show how it provides
a different way to look
at the prospects of unification in different cases. We then use these
variables to study both one and two step unification models. In addition
to providing, what we believe is a new way to test for unification, we
find two new results with interesting phenomenological implications which
to the best of our knowledge have not been discussed in the
literature.  
(i) The minimal supersymmetric left-right symmetric model
with the seesaw mechanism, which resisted unification in the four
dimenaional case can now be unified if the gauge fields are put in the
bulk and (ii) with non-canonical normalization of gauge couplings, we
find examples where the unification scale is around $10^{11}$ GeV or so
which has been advocated in a recent paper as the possible string scale
from various phenomenological considerations~\cite{quevedo}. 

The paper is organized as follows. In section two we present the basic ideas
that introduce the power law running; in section three we compare on
analytical and numerical basis these results with those obtained by the
implementation of the step by step approach which invoke the decoupling theorem
\cite{appel} at  each level of the Kaluza Klein tower. Next section is
devoted to discuss
the MSSM and the SM unification. Here our goal is to show that generically,
through the model independent analysis, the compactification scale is fixed by
the experimental accuracy in the gauge coupling constants. Moreover, in the
supersymmetric SU(5) theory the one loop prediction  for $\alpha_s$ could be
within  the experimental  value just by fixing the compactification scale close
below the usual unification mass, without the introduction of extra matter and
assuming that all the standard gauge and scalars  and perhaps the fermions
propagate into the bulk. Nevertheless,
the SU(5) unification makes this result unstable under two loop corrections. As
it has been pointed out in references~\cite{kaku,carone,quiros,frampton},
a way to
solve this problem is to modify the bulk content. We present a simple
choice where only the gauge bosons develop excited modes.  Section five is
dedicated to discuss the case of two steps models. Here,  based on the results
of the analysis, we  argue that  the excited modes of higher symmetries,
expected to be embedded in the unification theory, as the left right model for
instance,  may split the unification scale pushing such symmetries down the
compactification scale. Moreover, as the MSSM particles could be naturally 
accommodated in left right representations, we argue that the left right model
must appear  below the compactification scale, fixing the unification and
setting lower bounds to the compactification scale.  We shall also show that in
some scenarios,  this effect could produce consistent results with both the
neutrino physics~\cite{nphys,see-saw} and proton decay.

%%%%%%%%%%%%%%%%%%%%%%%%%%%%%%%%%%%%%%%%%%%%%
\section{Power law running}
%%%%%%%%%%%%%%%%%%%%%%%%%%%%%%%%%%%%%%%%%%%%%
The evolution of the gauge coupling constants above the compactification
scale  $\mu_0$ was derived by Dienes, Dudas and Gherghetta 
(DDG)~\cite{ddg1,ddg2}
on the base of an effective (4-dimensional) theory approach. 
The general result at one-loop level is given by
 \be 
 \alpha_i^{-1}(\mu_0) = \alpha_i^{-1}(\Lambda) + {b_i - \tilde{b}_i\over 2\pi}
 \ln\left( {\Lambda\over \mu_0}\right) + {\tilde{b}_i\over 4\pi}\ 
 \int_{r\Lambda^{-2}}^{r\mu_0^{-2}}\! {dt\over t} 
 \left\{ \vartheta_3\left({it\over \pi R^2}\right)\right\}^\delta , 
 \label{exact}
 \ee
with $\Lambda$ as the ultraviolet cut-off, $\delta$  the number of extra
dimensions and $R$ the compactification radius identified as $1/\mu_0$. The
Jacobi theta function
 \be
 \vartheta(\tau) = \sum_{n=-\infty}^{\infty} e^{i\pi \tau n^2}
 \ee
reflects the sum over the complete (infinite) Kaluza Klein (KK) tower. In Eq.
(\ref{exact}) $b_i$ are the beta functions of the theory below the $\mu_0$
scale, and $\tilde{b}_i$ are the contribution to the beta functions 
of the KK states at each excitation
level. Besides, the numerical factor $r$ in the former integral could not be
deduced purely from this approach. Indeed, it is obtained assuming that
$\Lambda\gg \mu_0$ and comparing the limit with the usual renormalization group
analysis, decoupling all the excited states with masses above $\Lambda$, and also
assuming that the number of KK states below certain energy $\mu$ between $\mu_0$
and $\Lambda$ is well approximated by the volume of a $\delta$-dimensional
sphere of radius $\mu/\mu_0$
 \be 
 N (\mu,\mu_0) = X_\delta
 \left({\mu\over\mu_0}\right)^\delta ; 
 \label{npl}
 \ee
with $X_\delta = \pi^{\delta /2}/\Gamma(1 +\delta /2)$.
The result is a power law behaviour of the gauge coupling constants given by
 \be
 \alpha_i^{-1}(\mu) = \alpha_i^{-1}(\mu_0) - {b_i - \tilde{b}_i\over 2\pi}
 \ln\left( {\mu\over \mu_0}\right) - {\tilde{b}_i\over 2\pi}\cdot
 {X_\delta\over\delta}\left[ \left({\mu\over\mu_0}\right)^\delta - 1\right] .
 \label{ddgpl}
 \ee

Nevertheless, as it was pointed out by Ghilencea and Ross~\cite{ross}, in the
MSSM the energy range between $\mu_0$ and $\Lambda$ --identified as the
unification scale-- is relatively small due to the steep behaviour in the
evolution of the couplings. For instance, for a single extra dimension the
ratio $\Lambda/\mu_0$ has an upper limit of the order of 30, which 
substantially decreases for higher $\delta$ to be less than 6.

Clearly, this fact seems in conflict with the assumption which justifies Eq.
(\ref{npl}).  Moreover, as the number of KK states with masses lesser than
$\mu$ are by definition the number of solutions to the equation
 \be
 \sum_{i=1}^\delta n_i^2 \leq \left({\mu\over \mu_0}\right)^2,
 \ee
where $n_i\in\IZ$ corresponds to the  component of the
 momentum of the fields propagating into the bulk along the $i$-th extra
 dimension; we expect
that only the first few levels are relevant in the analysis for $\mu\sim
\mu_0$, restoring the usual logarithmic behaviour rather than the power
behaviour as in Eq. (\ref{ddgpl}). It is then important to know the kind of
errors involved in the power law assumption.

%%%%%%%%%%%%%%%%%%%%%%%%%%%%%%%%%%%%%%%%%%%%%
\section{Logarithmic running}
%%%%%%%%%%%%%%%%%%%%%%%%%%%%%%%%%%%%%%%%%%%%%

In general, the mass of each KK mode is well approximated by 
 \be 
 \mu_n^2 = \mu_0^2\sum_{i=1}^\delta n_i^2. 
 \label{mn}
 \ee
Therefore, at each mass level $\mu_n$ there are as many  modes as
solutions to Eq. (\ref{mn}). 
It means, for instance, that in one extra dimension each
KK level will have 2 KK states that match each other, 
with the exception of the zero modes which are
not degenerate and correspond to (some of) the particles 
in the original (4-dimensional)
theory manifest below the $\mu_0$ scale. In this particular case, 
the mass levels are separated by units of $\mu_0$. 
In higher extra dimensions the KK levels are
not regularly spaced any more. Indeed, as it follows from Eq. (\ref{mn}), the
spectra of the excited mass levels correspond to that of the energy in the 
$\delta$-dimensional box, where the degeneracy of each energy level is not so
trivial but still computable (see below).

{\it One extra dimension.}
The one-loop  renormalization group equations for energies just above the
$n$-th level  ($\mu> n\mu_0$) in the simplest case, $\delta = 1$, are
 \be 
 {d\over d\ln\mu}\alpha_i^{-1} = -{b_i+ 2n \tilde{b}_i \over 2 \pi},
 \ee
for at this level all the low energy particles contribute through 
$b_i$ --which already 
include the zero modes-- and all the $2n$ excited
states in the first $n$ KK levels, each of one giving a contribution of
$\tilde{b}_i$. Solving this equation requires 
boundary conditions at $n\mu_0$, to get
 \be
 \alpha_i^{-1}(\mu) = \alpha_i^{-1}(n\mu_0) - {b_i +2 n \tilde{b}_i\over 2\pi}
 \ln\left( {\mu\over n\mu_0}\right),
 \ee
while, for the same arguments
 \be
 \alpha_i^{-1}(n\mu_0) = \alpha_i^{-1}\left((n-1)\mu_0\right) 
 - {b_i + 2 (n-1) \tilde{b}_i\over 2\pi} \ln\left( {n\over n-1}\right),
 \ee
and so on, up to
\be
 \alpha_i^{-1}(2\mu_0) = \alpha_i^{-1}\left(\mu_0\right) 
 - {b_i + 2  \tilde{b}_i\over 2\pi} \ln\left( 2\right).
 \ee
Combining all these equations together is straightforward to get
  \be
 \alpha_i^{-1}(\mu) = \alpha_i^{-1}(\mu_0) - {b_i\over 2\pi}
 \ln\left( {\mu\over \mu_0}\right) - {\tilde{b}_i\over 2\pi}\cdot 2\left[
 n\ln\left(\mu\over\mu_0\right) - \ln n! \right].
 \label{log}
 \ee
which explicitly shows a logarithmic behaviour just corrected by the
appearance of the $n$ thresholds below $\mu$. 

Using the Stirling's formula $n!\approx n^n e^{-n} \sqrt{2\pi n}$ valid for
large $n$,  the last expression takes the form of the power law running 
 \be
 \alpha_i^{-1}(\mu) = \alpha_i^{-1}(\mu_0) - {b_i-\tilde{b}_i\over 2\pi}
 \ln\left( {\mu\over \mu_0}\right) - {\tilde{b}_i\over 2\pi}\cdot 2\left[
 \left(\mu\over\mu_0\right) - \ln \sqrt{2\pi} \right].
 \label{limit}
 \ee
The last term $\ln\sqrt{2\pi}\approx 0.9189$, thus, the limit is fully
consistent with eq (\ref{ddgpl}). Indeed this small difference could be
absorbed by high energy threshold or even second order corrections.

It is worth pointing out that while in the limiting case of large number
of Kaluza-Klein states Eq. (12) agrees with Eq. (4), for the case of
finite N, there is a difference as can be seen by choosing $\mu =
\mu_0+\epsilon$ in Eq. (4) and Eq. (11).

{\it Higher dimensions.} For the case of two or more extra
dimensions, as in its classical
quantum analogous, each level is characterized by the set of natural numbers
$\{n_1,\dots,n_\delta\}$ which satisfy Eq. (\ref{mn}). It is clear that if
all the $n_i$ numbers are different and non zero, the KK level is $2^\delta
\delta !$-fold degenerated, since there are $\delta !$ ways of distributing
these $\delta$ (absolute) values between the $\delta$ numbers $n_i$, and there
are $2^\delta$ different combinations of the signs for each one of those
combinations. Besides, some of these numbers could be equal or even zero, then,
in general the degeneracy of each level is given by 
 \be
 g_N = 2^{\delta-p} {\delta !\over k_1! k_2!\cdots k_l! p!};
 \ee
where $k_i$ is the number of times that the value (without sign) of $n_i$
appears in the array $\{n_1,\dots,n_\delta\}$, and $p$ is the number of
zero elements in the same array. The (natural) 
index $N$ stands for the label of the level
corresponding to the squared ratio of masses
 \be
 \sum_{i=1}^\delta n_i^2 = \left({\mu_{N-1}\over \mu_0}\right)^2 \equiv N 
 \qquad ; \quad N \in \IN .
 \label{N}
 \ee
In addition to this ``normal'' degeneracy, there often are 
accidental degeneracies due to
certain numerical coincidences: some natural numbers have more than one
non equivalent decompositions of the form (\ref{N}). For instance, for
$\delta=2$, we can write $25= 5^2 + 0 = 4^2 + 3^2$, thus  level 25 is 12-fold
degenerated (4 times from the first decomposition plus 8 times from the second
one), while level 5 is just 8-fold degenerated ($5= 2^2 + 1$)
and level 3 does not exist.

Despite this complexity of the spectra, the degeneracy of each level 
is always computable and performing a level by level approach of the gauge
coupling running is still possible. In this case,
the renormalization group equations for
energies above the $N$-th level receive contributions from $b_i$ and of
all the KK excited states in the
levels below, in total
 \be
 f_\delta(N) = \sum_{n=1}^N g_\delta(n) ;
 \ee
where $g_\delta(n)$ represent the total degeneracy of the level $n$. The
coupling evolution equations then look like
 \be
 \alpha_i^{-1}(\mu) = \alpha_i^{-1}(\mu_{N-1}) - 
 {b_i + f_\delta(N) \tilde{b}_i\over 2\pi}
 \ln\left( {\mu\over \mu_{N-1}}\right).
 \ee
Iterating this result for all the first $N$ levels and combining all
them together with Eq. (\ref{N})
we get the logarithmic  running given by
 \be
 \alpha_i^{-1}(\mu) = \alpha_i^{-1}(\mu_0) - {b_i\over 2\pi}
 \ln\left( {\mu\over \mu_0}\right) - {\tilde{b}_i\over 2\pi} \left[
 f_\delta(N)\ln\left(\mu\over\mu_0\right) - 
 {1\over 2}\sum_{n=1}^N g_\delta(n)\ln n  \right],
 \label{hdlog}
 \ee
where now the correction of the $N$ thresholds appears a little bit more
complex than as before. It is clear that this relationship reduces to Eq.
(\ref{log}) for $\delta=1$.

For large $N$, where the Eq. (\ref{npl}) holds, the former expression
reduces to power law running as we show now.
 To prove this, let us take only the last
term in brackets in Eq. (\ref{hdlog}), which, using that $g_\delta(n) =
f_\delta(n) - f_\delta(n-1)$, may be rewritten in the form
 \bea
 F_\delta\left({\mu\over\mu_0}\right) &\equiv &
 f_\delta(N)\ln\left(\mu\over\mu_0\right) - 
 {1\over 2}\sum_{n=1}^N g_\delta(n)\ln n   \label{f} \\[1ex]
 & = & 
 f_\delta(N)\left[ \ln\left({\mu\over\mu_0}\right) - {1\over 2}\ln N\right]
 - {1\over 2} \left[ \sum_{n=1}^N\bigg(f_\delta(n) - f_\delta(n-1)\bigg)\ln n 
 - f_\delta(N) \ln N\right].
 \eea
In the limit when $N$ is large, $N\approx (\mu/\mu_0)^2$. Hence, the first term
vanishes. The remaining terms becomes in the continuum limit
 \be 
 F_\delta\left({\mu\over\mu_0}\right) = 
 -{1\over 2} \left[ \int_1^N  dn\  {d f_\delta \over dn}(n) \ln n 
 - f_\delta (N)\ln N \right ] = \int_{\mu_0}^\mu {d\mu'\over \mu'}
 f_\delta(n(\mu')) .
 \ee
Assuming in  this limit that $f_\delta(n(\mu))
\approx X_{\delta}\left( \frac{\mu}{\mu_0}\right)^\delta-1$, we recover
the
DDG approximation for large $N$:
 \be 
 F_\delta\left({\mu\over\mu_0}\right)  \approx
 {X_\delta\over\delta}\left[ \left({\mu\over\mu_0}\right)^\delta - 1\right]
  - \ln\left({\mu\over\mu_0}\right) .
 \ee

The explicit difference in the running of the gauge coupling constants governed
by the power law (\ref{ddgpl}) from the logarithmic running  (\ref{hdlog}) is
best appreciated from the numerical analysis of the function $F_\delta$, since
$[\alpha_i^{-1}(\mu)]^{DDG} - [\alpha_i^{-1}(\mu)] =  \tilde{b}_i
[F_\delta (\mu/\mu_0) - F_{\delta}^{DDG}(\mu/\mu_0)]/{2\pi}$;  where
the index $DDG$ stands to identify the power law expressions. Such difference 
has been plotted in figure 1 for one and two extra dimensions. We
notice  for $\delta=1$ that $F_\delta- F_\delta^{DDG}$ tends to
converge quickly to an asymptotic  value, while, for $\delta=2$ it is more
unstable but still convergent. This fact is a consequence of the complexity in
the level degeneracy. From here, is easy to figure out that a more unstable
behaviour arise for higher  $\delta$. Indeed, for $\delta=3$ we found that a
more large number of thresholds is required to stabilize the difference into a
small slope of the asymptotic value, which tends to be higher for larger extra
dimensions.

Nevertheless, as $F_\delta$ has a steep evolution, in the limit of large
$N$ those differences becomes to be strongly suppressed compared with the actual
value of $F_\delta$, which dominates the running of 
the gauge coupling constants,
as it is depicted in figure 2, where we can observe that for $\mu/\mu_0>10$, 
$F_\delta- F_\delta^{DDG}$ represent only less than 2\% of the value of
$F_\delta$. However, for lower ratios  the deviation of 
$F_\delta$ from the power law remains within 2\% to 50\%.

Before closing the present section, we point out a natural
extension to our analysis to the case where the $\delta$ compactification
radii $R_i$ are not all equal, since
the requirement of their equality is so far unjustified.  In this case,
the masses of the excited KK modes are given by
 \be
 m_n^2 = \sum_{i=1}^\delta n_i^2\mu_i^2;
\label{mmn}
 \ee
where we have defined $\mu_i=1/R_i$. Assuming that $\mu_0= 1/R_{max}$, the
inverse of the largest radius, the
contributions of the bulk to the renormalization group equations should start
at these mass levels. Hereafter the running will cross a new threshold
each
time that $\mu$ reachs a level in the tower, again characterized, as those of
the non cubic $\delta$ dimensional box, by the squared ratio of masses
 \be
 M_n \equiv \left({m_n\over \mu_0}\right)^2 =
 \sum_{i=1}^\delta n_i^2 \left({\mu_i\over \mu_0}\right)^2 .
 \ee
As it is clear, our approach will work following the same steps as before, and
then, we will arise to  a logarithmic running of the same form as in Eq.
(\ref{hdlog}), but now with $F_\delta$ given by
 \be
 F_\delta(\mu,\mu_0,\cdots\mu_\delta) = 
 f_\delta(N)\ln\left(\mu\over\mu_0\right) - 
 {1\over 2}\sum_{n=1}^N g_\delta(n)\ln M_n ,
 \ee
for $\mu$ just above the $N$-th level of the tower. In the continuous limit (when
$\mu$ is large compared with whatever $\mu_i$) we may
assume that the number of states below the energy scale $\mu$ is well
approximated by the volume of the  $\delta$ dimensional ellipsoid defined by Eq.
(\ref{mmn}) where $m_n\approx \mu$
 \be
 N(\mu,\mu_0,\cdots\mu_\delta)\approx 
 X_\delta \prod_{i=1}^\delta\left(\mu\over\mu_i\right).
 \ee
In this limit
 \be 
  F_\delta(\mu,\mu_0,\cdots\mu_\delta)  \approx
 {X_\delta\over\delta}\left[ \prod_{i=1}^\delta\left({\mu\over\mu_i}\right)
  - \prod_{i=1}^\delta\left({\mu_0\over\mu_i}\right) \right]
  - \ln\left({\mu\over\mu_0}\right) .
  \label{fdeg}
 \ee
with the explicit extraction of the zero modes. Clearly, when all the radii are
equal it reproduces the DDG expression.

%%%%%%%%%%%%%%%%%%%%%%%%%%%%%%%%%%%%%%%%%%%%%
\section{Unification in extra dimensions}
%%%%%%%%%%%%%%%%%%%%%%%%%%%%%%%%%%%%%%%%%%%%%

Let us now analyze the implications of extra dimensions for unification
of the gauge couplings. Many features of unification can be studied
without bothering about the detailed subtleties concerning the logarithmic
vrs power law running. So we will simply use the generic form for the 
evolution equation suggested by Eq. (17) i.e.
 \be
 \alpha_i^{-1}(M_Z) = \alpha^{-1} + {b_i\over 2\pi}
 \ln \left( {\Lambda\over M_Z}\right) + {\tilde{b}_i\over 2\pi} 
 F_\delta\left( {\Lambda\over \mu_0}\right), 
 \label{rgef}
 \ee
being $\alpha$ the unified coupling.
It is clear from Eq. (27) that the information that comes from the bulk
can be separated into
two independent parts: all the structure of the spectra of the KK tower,
defined by the compactification scale and the number of extra dimensions is
completely embedded into the $F_\delta$ function, in such a way that its
contribution to the running of the gauge couplings is actually model
independent. The model dependence coming from such questions as to which
representations are in bulk etc are all encoded in the beta functions
$\tilde{b}_i$.
In other words, no matter how the fields of the low energy four dimensional
theory are distributed among the bulk and the wall, the $F_\delta$ function is
not affected and conversely, changes in the KK tower, namely the splitting
in the
compactification radii, affect only the form of $F_\delta$ through the
changes in the internal mass spectra [Eq.(\ref{fdeg})].

Notice that Eq. (\ref{rgef}) is similar to that of  the two step unification
model, where a new gauge symmetry appears at an intermediate energy scale.
It was already noted sometime ago~\cite{mohapatra} that
the solutions to the renormalization group equations in those models are
very constrained by the one step unification  in the MSSM. The
argument goes as follows: let us define the vectors 
${\bf b}= (b_1,b_2,b_3)$; 
$\tilde{\bf b}= (\tilde{b}_1,\tilde{b}_2,\tilde{b}_3)$; 
${\bf a} = (\alpha_1^{-1}(M_Z),\alpha_2^{-1}(M_Z),\alpha_3^{-1}(M_Z))$ and 
${\bf u} = (1,1,1)$ and
note that Eq. (\ref{rgef}) takes the simplest vectorial form
 \be
 {\bf a} = \alpha^{-1} {\bf u}  + x {\bf b} + y \tilde{\bf b}
 \label{rgev}
 \ee
where $x = \ln(\Lambda/M_Z)/2\pi$ and $y = F_\delta/2 \pi$. From here is
easy to eliminate  variables, for instance $\alpha$ and $x$
to get
 \be
 \Delta\alpha\equiv ({\bf u}\times {\bf b})\cdot {\bf a} = 
 [({\bf u}\times {\bf b})\cdot \tilde{\bf b}]\ y.
 \label{dalpha}
 \ee
In the MSSM ${\bf b} = (33/5,1,-3)$. Now, inserting the experimental
values~\cite{pdg}
 \bea
 \alpha^{-1}_1(M_Z) &=& {3\over 5}\alpha^{-1}_Y(M_Z) 
  = 58.9946 \pm 0.0546, \nonumber\\
 \alpha^{-1}_2(M_Z) &=& 29.571\pm 0.043, \quad\mbox{and} \label{alphas}\\
 \alpha^{-1}_3(M_Z) &=& \alpha_s^{-1}(M_Z) = 8.396\pm 0.127. \nonumber
 \eea
we get $\Delta\alpha_{expt} = 0.928\pm 0.517$. So, $\Delta\alpha_{expt}$
is consistent
with zero within two standard deviations. One step unification condition
is of course given by $\Delta\alpha=0$. Thus, assuming one step
unification, we find that the Eq. (\ref{dalpha}) reduces to the
constraint~\cite{mohapatra}
 \be
 (7\tilde{b}_3 - 12\tilde{b}_2 + 5 \tilde{b}_1)F_\delta= 0;
 \ee
since $[({\bf u}\times {\bf b})\cdot \tilde{\bf b}] = -4(7\tilde{b}_3 -
12\tilde{b}_2 + 5 \tilde{b}_1)/5$. 
This equation was first written down in Ref.~\cite{mohapatra} for the case
of generic unification with an intermediate scale and has subsequently
been re-used in several recent papers\cite{quiros,frampton}.
There are two solutions to the last equation: 
\bi
\item[(a)] $F_\delta(\Lambda/\mu_0)=0$, which means $\Lambda = \mu_0$,
bringing  us back to the MSSM by pushing up the compactification scale to
the unification scale.
\item[(b)] Assume that the beta coefficients $\tilde{b}$ conspire to eliminate
the term between brackets:
 \be 
 (7\tilde{b}_3 - 12\tilde{b}_2 + 5 \tilde{b}_1) = 0 .
 \label{cons}
 \ee
The immediate consequence of Eq. (\ref{cons}) is the indeterminacy  of
$F_\delta$, which means that we may put $\mu_0$ as a free parameter in the
theory. For instance we could choose $\mu_0\sim$ 10 TeV to maximize the
phenomenological impact of such models~\cite{pheno}. 
It is compelling to stress that
this conclusion is independent of the explicit form of $F_\delta$.
\ei

Furthermore, as 
 $ [({\bf u}\times {\bf b})\cdot \tilde{\bf b}] = 
 (b_1- b_3)(\tilde{b}_2- \tilde{b}_3) - (b_2- b_3)(\tilde{b}_1- \tilde{b}_3);
 $
Eq. (\ref{cons}) is equivalent to the condition obtained by DDG expressed by 
 \be
 {B_{12}\over B_{13} } = {B_{13}\over B_{23} } = 1; \qquad \mbox{where}\qquad
 B_{ij} = {\tilde{b}_i- \tilde{b}_j\over b_i- b_j} .
 \ee
 	
What is clear at this point, is that the DDG minimal model  can not satisfy Eq.
(\ref{cons}) by itself, since it contains the MSSM fields plus extra fermionic
and scalar matter not matching that of the MSSM~\cite{ddg2}, for the fields in
the bulk are accommodated in $N=2$ hyperrepresentations, implying a higher
prediction for $\alpha_s$ at low $\mu_0$~\cite{ross}. Indeed, in this case we
have $(7\tilde{b}_3 - 12\tilde{b}_2 + 5 \tilde{b}_1) = -3$. However,  as
Carone~\cite{carone} showed, there are some  scenarios where option (b) may be
realized. In those cases, the MSSM fields are distributed in a nontrivial 
way among
the bulk and the boundaries. There are also other possible extensions to those
scenarios. We may add matter to the MSSM,  as long as we do not affect the
constraint (\ref{cons}), in order to get the same  MSSM accuracy for
$\alpha_s$. Also, bulk fields with non zero modes could be added to the scheme
to satisfy Eq. (\ref{cons}). Those cases have been considered in
references~\cite{kaku,quiros,frampton}. 

In fact looked from this perspective, the embedding of the standard model
into higher dimensional space-time looks worse. The reason is that
since ${\bf b}^{SM} = (41/10,-19/6,-7)$ we have
$\Delta\alpha_{expt}=41.13\pm 0.655\neq 0$;
as is well known this is the reason for the failure of grand unification
of the standard model. Now let us form a
second linear combination obtained from (\ref{rgev}), namely
 \be
 \tilde{\Delta}\alpha\equiv ({\bf u}\times \tilde{\bf b})\cdot {\bf a} = 
 [({\bf u}\times \tilde{\bf b})\cdot {\bf b}]\ x  =
  - [({\bf u}\times {\bf b})\cdot \tilde{\bf b}]\ x .
 \label{delta2}
 \ee
Eq. (34) implies that in order to get a good solution where $x$ and $y$
are positive (as they are by definition since
 $\Lambda >M_Z$ and $\Lambda >\mu_0$), the following
constraint must be satisfied
 \be 
 Sign(\Delta\alpha)  
 = Sign[({\bf u}\times {\bf b})\cdot \tilde{\bf b}]
 = - Sign(\tilde{\Delta}\alpha) .
 \label{cons2}
 \ee
However, in the minimal model where all fields are assumed to have  KK
excitations $\tilde{\bf b} = (1/10,-41/6,-21/2) + (4/3,4/3,4/3)\eta$, with
$\eta$ the number of generation in the bulk, we get 
$\tilde{\Delta}\alpha = 38.973\pm 0.625$; and $({\bf u}\times {\bf b}^{SM})\cdot
\tilde{\bf b}^{SM} = 1/15$. Hence, the constraint (\ref{cons2}) is not
fulfilled and unification does not occur. Moreover, these results mean
that $\Lambda< M_Z$, which clearly is an inconsistent solution. However,
extra matter could of course improve this 
situation~\cite{quiros,frampton}.

		%%%%%%%%%%%%%%%%%%%%%%%%%%%%%%%
		
Note further that strictly speaking $\Delta\alpha_{expt}\neq 0$ in the
MSSM within the
experimental accuracy of one standard deviation and thus the right
condition that must apply over
and above Eq. (\ref{cons2}) is
 \be
 \alpha^{-1} =  
 { ({\bf b}\times \tilde{\bf b})\cdot {\bf a} \over
   ({\bf u}\times {\bf b})\cdot \tilde{\bf b} } >1 
   \label{cons3}
 \ee
obtained also from Eq. (\ref{rgev}), which insures that $\alpha$ remains in
the perturbative regime. Therefore, whenever the conditions (\ref{cons2}) and
(\ref{cons3}) are satisfied by $\tilde{\bf b}$, a unique solution for
$\Lambda$, and $\mu_0$ can be obtained from eqs. (\ref{dalpha}) and
(\ref{delta2}).  

Let us now turn to the cases where
the normalization of the gauge coupling differs from that of SU(5)
introduced above. Indeed for a gauge group  $G$ with coupling constant
$\alpha$, where the simple group $G_i$ is embedded, the coupling constant
$\alpha_i$ of $G_i$ is related to $\alpha$ by a linear relationship 
$\alpha_i = c_i \alpha$, with the (embedding) factor  defined by
 \be 
 c_i = {Tr\ T^2 \over Tr\ T_i^2} ;
 \ee
where $T$ is the generator of $G_i$ normalized over a representation $r$ of
$G$, and $T_i$ is the same generator but normalized over the representations
of $G_i$ contained into $r$; the traces run over the  complete representation
$r$~\cite{ponce}. 

In Table I we present the embedding $c_i$  factors  for several models of
general interest as they were introduced in references~\cite{models}. The
first
entry in that table correspond to most of the models in the literature. In 
models with $c_3 = {1\over 2}$  the color group SU(3) is embedded through the 
chiral color  extension~\cite{2su3} SU(3)$_{cL}\times$SU(3)$_{cR}$. In general 
$c_2 = 1/F$, where $F$ is the number of families contained in the same
representation~\cite{comment}. $c_1$ depends in how the hypercharge is embedded
into the group and is given by $c_1^{-1} = {1\over 2} Tr (\frac{Y}{2})^2$.
In SU(5), for
instance,  $(c_1,c_2,c_3)= ({3\over 5},1,1)$. These are the canonical values.
They reflect the  absence of extra fields in the fermionic representations with
non trivial quantum numbers.    The group $[SU(4)]^3\times Z_3$  in Table I is
not the vector-like color version of  the two family $SU(4)_{color}$ 
model, but it
is the one family model introduced in reference~\cite{models}. Last two models
in Table I are three family $SU(4)_{color}$~\cite{patis} models.
We are considering these models as
examples of semisimple unified theories.

The evolution equations are still given by (\ref{rgev}), but with the
$c_i$ factors included in both the gauge coupling constants, 
${\bf a} = (c_1\alpha_Y^{-1}(M_Z), 
c_2 \alpha_2^{-1}(M_Z),c_3 \alpha_3^{-1}(M_Z))$  
and beta functions
${\bf b}= (c_1 b_Y,c_2 b_2,c_3 b_3)$; 
$\tilde{\bf b}= (c_1\tilde{b}_Y,c_2 \tilde{b}_2,c_3 \tilde{b}_3)$. The
unnormalized beta functions for the MSSM are 
$(b_Y,b_2,b_3) = (11,1,-3)$ and  for the SM 
$(b^{SM}_Y,b^{SM}_2,b^{SM}_3) = (41/6,-19/6,-7)$. Using these values and
the experimental data (\ref{alphas}), we find that $\Delta\alpha\neq 0 $
for all those models (Table I also).  The reason is that in non canonical
models extra scalars at the $M_Z$ scale are required in order to achieve
unification in both the MSSM and SM~\cite{ponce}.

Now, we proceed as follows. We first assume all gauge bosons as high dimensional
fields, then, for models in Table I we explore the scenarios
where the other MSSM (SM) fields propagate into the bulk: Higgs and/or fermion
fields.
 In the analysis we always
consider all the three families together. The contribution to the unnormalized
$\tilde{b}_i$ from these particles are given in Table II and they have been
calculated assuming that mirror fields are contained in all chiral $N=2$
hypermultiplets of the KK tower.  For those scenarios that fulfill the
conditions above we calculate $\Lambda$, $\mu_0$ and $\alpha$ using both, the
power law and the logarithmic approaches for $\delta=1$  using the central
experimental  values of the gauge coupling constants  and $M_Z=91.186$ GeV
as inputs. The results are shown in Table III. We have not found solutions for
the SM with this minimal content.

What is important to emphasize from these results is that in the supersymmetric
SU(5) class of models (all models with canonical normalization),  the
scenario where all the
fields live in the bulk
predicts a unification scale at $10^{16}$ GeV and a compactification
scale is slightly below this scale. Here the small $\Lambda/\mu_0$ ratio
makes it important to consider logarithmic approach rather than the power
law. As a matter of fact, only the first one or two  levels of the tower
contributes to the renormalization group equations. Thus,
it means that at the one loop level only few thresholds near the
unification scale are required to predict the right value for
$\alpha_s\approx 0.119$, bringing single scale MSSM models into better
agreement with experiments.

Comparing Table III with previous results for the SM (see
references~\cite{quiros,frampton}),
we may note that the scenarios where a solution
could be obtained are very different for the MSSM than for the SM (whether
we choose canonical or noncanonical normalization). 
The reason is again  that the SM by itself require in
general a large number of extra chiral scalars  or fermions to get
unified~\cite{ponce}. Part of
these kind of fields are provided by supersymmetry,
but as it follows from Table I, only in the canonical class do the new
fields of supersymmetry bring $\Delta\alpha$ close to zero.
When SM is embedded into higher dimensions, a new class of
contributions corresponding to excited scalar or fermionic
modes of the KK tower emerge\cite{frampton} and without any need for
supersymmetry, they lead to unification of gauge couplings. It must be
pointed out however that in most cases extra chiral fields carrying color
quantum numbers make $\alpha$ run into the non perturbative regime
of the theory. On the other hand, in the MSSM, extra scalars and fermions are
contained in the hypermultiplets of the KK modes; since there are more
particles contributing to beta functions, their contributions should be
pushed up to high mass scales in order to preserve unification.

	%%%%%%%%%%%%%%%%%%%%%%%%%%%%%%%%%%%%%%%%%%%%%

Let us now comment on the effect of the two loop contributions to the
coupling evolution. Since in this case the evolution equations change, in
order to use our method to understand the results for unification, we
redefine the values of $\alpha^{-1}_i(M_Z)$ by moving the two loop
contributions to the left hand side of the evolution equations. 
Now, since two loop corrections tend to increase the one loop predictions for
$\alpha_s$, once we subtract them from $\alpha^{-1}_s(M_Z)_{expt}$, it
tends to make $\Delta\alpha$ negative in the case of MSSM. This upsets
the consistency of Eq. (\ref{cons2}) and makes the possibility of
extra dimensions below the GUT scale less likely~\cite{ross}.
This can be avoided if one changes the scenario by including  extra
particles or by picking up some of those in the
bulk~\cite{kaku,carone,quiros,frampton}. Along this line, a
simple possibility is to let the gauge bosons propagate in the bulk,
a case which seems not to have been considered so far.
Our finding in this case including two loop corrections is that
 we get the predictions: $\Lambda\sim
1.282\times 10^{14}$ GeV; $\mu_0\sim 2.4 \times 10^{13}$ GeV and $\alpha^{-1} =
29.6122$.

On the other hand, notice that  the SM  and supersymmetric  non canonical
models ( such as SU(5)$\times$SU(5)) are not so sensitive to the
inclusion of the two loop effects, since
$\Delta\alpha$ starts out as a large positive number and retains its sign
when two loop effects are subtracted as indicated above.
This has the implication that our predictions only  will change
slightly when two loop corrections are included. A particularly
interesting outcome in the case of models with noncanonical is that
there are several cases where, the unification scale comes out to be
around $10^{11}$ GeV ( see Table III, cases $SU(5)\times SU(5)$,
$[SU(3)]^4$, $[SU(6)]^4$ ). These models fit nicely into the new
intermediate string scale models recently proposed in ~\cite{quevedo}.
Turning this point around, one could presume that the preferred GUT
group in the case of such string models would be the ones such as
$[SU(5)]^2$ etc rather than the canonical SO(10) or SU(5). It is worth
noting that such models have a number of phenomenologically desirable
features such as automatic R-parity conservation, no baryon violation
from the gauge theory etc. In fact, it is the property of automatic
baryon number conservation that makes such low unification scales
phenomenologically acceptable.

%%%%%%%%%%%%%%%%%%%%%%%%%%%%%%%%%%%%%%%%%%%%%
\section{Unification in two step models with extra dimensions}
%%%%%%%%%%%%%%%%%%%%%%%%%%%%%%%%%%%%%%%%%%%%%

Two steps unification models are of great current interest mainly motivated by
neutrino physics~\cite{nphys,see-saw}. In them
the general picture is as follows: at the $M_Z$ scale we have the MSSM
theory,
which remains valid up to certain intermediate scale $M_I$. Hereafter a new
gauge symmetry  $G'$  rules the evolution of the gauge couplings
up to the unification scale $M$. In this framework the one loop
renormalization group equations are given by
 \be
 \alpha_i^{-1}(M_Z) = \alpha^{-1}_M + {b_i\over 2\pi}
 \ln \left( {M_I\over M_Z}\right) + {b'_i\over 2\pi} 
 \ln\left( {M\over M_I}\right), 
 \label{rg2s}
 \ee
where $\alpha_M$ is the unified coupling and $b'_i$ the beta functions of the
$G'$ theory. This equation resembles the case discussed in the previous
section, and we will therefore try the same procedure to solve it.
In terms of $\Delta'\alpha\equiv ({\bf u}\times{\bf b}')\cdot {\bf a}$,
the condition
to get a good solution where the hierarchy $M>M_I>M_Z$ is  fulfilled now
reads
  \be 
 Sign(\Delta\alpha)  
 = Sign[({\bf u}\times {\bf b})\cdot {\bf b}']
 = - Sign(\Delta'\alpha) .
 \label{con2s}
 \ee
Again, the unification in the MSSM (canonical models), will imply that 
$M_I\sim M$ unless  $(7b'_3 - 12b'_2 + 5 b'_1) \approx 0$. 
Some examples where the value of $\alpha_s$ was used to make the
$\Delta \alpha_{expt}$ nonzero and thus realize this scenario 
were presented in
ref.~\cite{mohapatra} in the context of the left right model (LRM)~\cite{lrm}.
Other possible ways to realize intermediate scales would be to add extra
scalars at the weak scale to MSSM, in which case the vector ${\bf b}$
changes again making $\Delta \alpha_{expt}\neq 0$ and thereby opening up a
way to have an intermediate scale; another way would be to consider 
non canonical models, where the change in the normalization again leads to 
$\Delta \alpha_{expt}\neq 0$ allowing now the intermediate scale~\cite{ponce2}.

Let us suppose that KK modes of the $G'$ theory appear at certain scale
$\mu_0$
below the unification scale. Once the gauge couplings cross $\mu_0$, the
steepness of the running  will imply the change of the unification scale, and
in the worst case the loss of unification. To fix this problem the
intermediate scale may be moved to a proper value $M_I'$ to restore 
the unification at a new
scale $\Lambda$. In the presence of extra dimensions, the
renormalization group equations are written as
 \be
 \alpha_i^{-1}(M_Z) = \alpha^{-1}_\Lambda + {b_i\over 2\pi}
 \ln \left( {M'_I\over M_Z}\right) + {b'_i\over 2\pi} 
 \ln\left( {\Lambda\over M'_I}\right) + 
 {\tilde{b}'_i\over 2\pi} F_\delta\left( {\Lambda\over \mu_0}\right).
 \label{rg3s}
 \ee
where  $\alpha_\Lambda$ is the new  unified coupling and $\tilde{b}'$ the beta
function of the excited modes. Now,  to understand the role of $\mu_0$,
 we proceed as before by defining the
vectors ${\bf u}= (1,1,1)$; ${\bf b} = (b_1,b_2,b_3)$;
${\bf b}'=(b'_1,b'_2,b'_3)$; 
$\tilde{\bf b}'=(\tilde{b}'_1,\tilde{b}'_2,\tilde{b}'_3)$.
It then follows from Eq. (40) that the following consistency condition
must be satisfied:
 \be
 \ln\left( {M'_I\over \Lambda}\right) = 
 \left[ {({\bf u}\times {\bf b} )\cdot \tilde{\bf b}' \over
         ({\bf u}\times {\bf b})\cdot {\bf b}' }\right]\
 F_\delta\left( {\Lambda\over \mu_0}\right)
 \label{split}
 \ee 
Note that the condition for the existence of an intermediate scale
in the presence of extra dimensions is not as strigent as it is for the
case without them. In the latter case, we must have vanishing of
$({\bf u}\times {\bf b})\cdot {\bf b}'$ and this normally means very
precise cancellation among the beta function coefficients that occurs
only rarely. And cases where
an intermediate scale could not occur before can now support such scales.
However, in order to get the right hierarchy $\Lambda > M'_I$ the bulk fields
must then satisfy
 \be
 Sign[({\bf u}\times {\bf b} )\cdot \tilde{\bf b}'] = 
 - Sign[({\bf u}\times {\bf b})\cdot {\bf b}'].
 \label{consplit}
 \ee

Lets assume now the left right model  
$G_{LR}\equiv SU(3)_c\times SU(2)_L\times
SU(2)_R\times U(1)_{B-L}$ as the intermediate theory. Let us also assume
as the content of the Higgs sector above the $M_I$ scale $n_T$ right
handed pairs of
triplets  $\Delta_R(1,1,3,-2)+ \bar{\Delta}_R(1,1,3,2)$, $n_B$  bidoublets 
$\phi(1,2,2,0)$ and perhaps $n_L$ pairs of left handed doublets 
$\chi_L(1,2,1,1) + \bar\chi_L(1,2,1,-1)$ and $n_R$ pairs of right handed
doublets $\chi_R(1,1,2,1) + \bar\chi_R(1,1,2,-1)$. With this content  
 \[ 
 {\bf b}'=\left({12\over 5},0,-3\right)+
 \left({3\over 5}(10 n_T + 2n_R + n_B+n_L),n_B+n_L,0\right),
 \] 
and thus
 \be
 ({\bf u}\times {\bf b})\cdot {\bf b}' = 
 {12\over 5} (3 (n_B+n_L)-10 n_T - 2 n_R + 3).
 \label{crosslr}
 \ee
In the simplest cases where $n_B=1,2$ and  $n_T=1$ without doublets,  which are
actually the minimal and next to minimal scenarios in the supersymmetric
version of the LRM~\cite{susylr} where the see-saw~\cite{see-saw} mechanism is
naturally implemented, we have  $({\bf u}\times {\bf b})\cdot {\bf b}'<0$,  
and  $\Delta'\alpha>0$; respectively. Eventually,  within the
experimental
accuracy this result means that at the one loop order a wrong hierarchy
$M_I>M>M_Z$ obtains.
Clearly this problem could be fixed by adding more scalars,  but also two loop
corrections may fix this problem in a natural way, even though we will still
have $M_I\sim M$. What is important to emphasize here, is that  the
condition (\ref{consplit})  now means that  
$({\bf u}\times {\bf b})\cdot \tilde{\bf b}'$ must be positive.

Now we procced to consider the possible contents of the bulk. The contribution
of the  LRM  $N=2$ hyperrepresentations to the  beta functions   $\tilde{\bf
b}'$ and to Eq. (\ref{split}) are shown in Table V, assuming mirror particles
in all fermionic and scalar representations. From the same table we can note
that there are several scenarios with the right positive  sign to change the
two step prediction of the model.  As an interesting example we have considered
the case where only all  the gauge bosons propagate into the bulk for the
minimal and next to minimal case with one extra dimension and plotted in figure
3 the splitting effect produced for these  KK modes over the unification scale 
as a function of the ratio $\Lambda/\mu_0$ by using the logarithmic approach. 
In order to stress this effect we have assumed for the plots that
$\Delta\alpha=0$, meaning the MSSM accuracy on $\alpha_s$.  The correction
introduced  by assuming the experimental accuracy of $\alpha_s$   only
will
produce an initial splitting,  as we argue above. However, the total behaviour
and our conclusions still remain.

In Fig. 4 and 5, we plot the running of the gauge couplings in this model
with one extra dimension and with only gauge bosons propagating in the
bulk. We find that the couplings unify with values of $M_I=M_{W_R}$
anywhere
from a TeV to $10^{16}$ GeV. It is however important to realize that the
values of the unification scale depends on $M_{W_R}$. If we want the
unification group to be SO(10), constraints of proton life time would
require that $\Lambda \geq 10^{15}$ GeV. This requires as we see from 
Fig. 4 that $M_{W_R}\simeq 10^{13}$ GeV (for the case of two bidoublets).
This value of the intermediate scale is of course what is required for
understanding the neutrino oscillation phenomena. The low value of 
$\Lambda$ also means that $p\rightarrow e^+ +\pi^0$ mode mediated via
the gauge boson exchange is now in the range accessible to the
super-Kamiokande experiment\cite{superK}.

As the value of the $M_{W_R}$ is lowered, the unification scale also goes
down. In these cases, considerations of proton decay would suggest that
the GUT group be some group other than SO(10) and furthermore string
related discrete symmetries be invoked to prevent catastrophic proton
decay.

  It is worth mentioning that  these
result do not change if we add the three families of fermions to the
bulk; however, the addition of the Higgs fields may produce the wrong sign
and then, this scenario
will split the unification scale with an inconsistent hierarchy. 

There is another  interesting case where the splitting is produced with the
right hierarchy. Lets assume that only the MSSM fields develop KK modes
(the gauge, Higgs and the matter).
If this is the case, the appearance of extra dimensions (i.e. $\mu_0\ll 
\Lambda$), below the unification scale pushes down the left right scale
to intermediate energies. This is clear from Figure 3. It is then clear
that if we accept that the left right scale has a lower bound about
800 GeV~\cite{pdg2},  this does impose a lower bound on the
compactification
scale $\mu_0$ about $10^{11}$ GeV. Moreover,  this bound seems to be 
independent of the bulk content since we obtain similar results for the cases
considered above. Other less trivial
scenarios that mix the LRM fields may produce a similar splitting but we
do not consider them here. 

From figure 3 we also note that  for $\delta=1$  only the first few 
thresholds are
required to get the right hierarchy $M_Z<M_I<\mu_0<\Lambda$.  

In contrast with the analysis performed in the previous section, now the two
loop contributions  will not affect our conclusions since so far they will
only  add an extra contribution to the splitting on the right way.

On the other hand,  as in  non canonical models and in those which add extra
matter to the MSSM we may expect to  start with an initial and may be more
important  splitting  as a consequence of the absence of one step unification,
the condition over the bulk matter (\ref{consplit}) will not apply any more,
unless the specific two step model predicts an initial wrong hierarchy, as it
really happens in the minimal LRM as discussed above.  Otherwise, the bulk
content is not constrained and the contributions of the KK modes will only
change the initial hierarchy. However, as we expect
that $M_I$ always be  smaller than $\mu_0$, whatever be the change, this
phenomenological condition will introduce a lower bound to the compactification
scale, although this bound will be very model dependent.

Finally, we note that keeping $\alpha_\Lambda$ in the
perturbative range also puts some constraints on the models as can be seen
from the following equation:
 \be
 \alpha_\Lambda^{-1} = \alpha_M^{-1} - 
  \left[ {({\bf b}\times {\bf b}' )\cdot \tilde{\bf b}' \over
         ({\bf u}\times {\bf b})\cdot {\bf b}' }\right]\
 F_\delta\left( {\Lambda\over \mu_0}\right) .
 \ee
Now, if the coefficient of $F_\delta$ is negative, $\alpha_\Lambda$  will
becomes smaller and the theory shall remains perturbative. This  is actually 
the case in the scenarios presented above, as long as fermions do not develop
KK modes. Otherwise, $\alpha_\Lambda$ will increase very quickly, following the
steepness of $F_\delta$, and  for small ratios of $\Lambda/\mu_0$ it will
cross to the non perturbative regime.

%%%%%%%%%%%%%%%%%%%%%%%%%%%%%%%%%%%%%%%%%%%%%
\section{Conclusion}
%%%%%%%%%%%%%%%%%%%%%%%%%%%%%%%%%%%%%%%%%%%%%

In conclusion, we have compared the logarithmic running with power law
running in theories with higher dimension and pointed out by explicit
examples that for realistic cases where, only few KK modes contribute,
the power law running may not accurately reflect the correct situation to
a high precision.  

We then analyze the general conditions to  achieve unification in 
presense of extra dimensions. We find that [Eq.(\ref{cons2})] provides
a generic constraint on the nature of the bulk fields for the
compactification scale to be below the unification scale.
We have considered both the cases where the low energy group is embedded
into the GUT group in a canonical or noncanonical manner and derive the
unification as well as the compactification scales from LEP data on the
low energy couplings. We find that in
supersymmetric SU(5) and other canonical models, the experimental accuracy at
one loop level requires  only the  first excited modes  of the MSSM fields to
appear below the unification scale. However, this model will predict
$\alpha_s$  at the MSSM accuracy at two loop order.  A new scenario not
considered before in the literature is the case where only the gauge
bosons propagate into the bulk. In this case the one loop order prediction for
$\alpha_s$ is small enough and is such that the two loop effects push it 
within the experimental error with an unification scale $\Lambda\sim
10^{14}$ GeV and $\mu_0\sim 10^{13}$ GeV.

Finally, considering the effect of extra dimensions on the two step models, we
found that in those models with the MSSM as the low energy limit, 
the presence of the KK excited modes makes it easier to have an
intermediate scale. We derive the constraint that
the models of this type must satisfy. We provide examples where, the
existence of an
$M_I$ depends on the bulk content. On the other hand,
if  only the MSSM fields develop excited modes, as it was assumed
in the DDG minimal model, their contributions can split the unification
scale, setting the supersymmetric left right model at intermediate energies.
We have found several examples of models where we have the
ordering of scales as $M_{W_R} \leq \mu_0 \leq \Lambda $. A particular
scenario where LRM gauge bosons are the only particles propagating in the
bulk (besides the graviton), we obtain the $W_R$ scale at $10^{13}$ GeV
which is of great interest for neutrino masses. Since the unification
scale in this case is of order $10^{15}$ GeV, it predicts proton life time
accessible to the ongoing Super-Kamiokande experiment.

We have also found some scenarios, where it is possible to get a lower
bound on the scale of the extra dimensions from two step unification and
present experimental lower bound on the $W_R$ scale.

%%%%%%%%%%%%%%%%%%%%%%%%%%%%%%%%%%%%%%%%%
\section*{Acknowledgments.}
%%%%%%%%%%%%%%%%%%%%%%%%%%%%%%%%%%%%%%%%%
 APL would like to thank the kind hospitality of the members of the Department
of Physics at U. MD. The work of APL is supported in part by CONACyT
(M\'exico). The work of RNM is supported by a grant from the National
Science Foundation under grant number PHY-9802551.

%%%%%%%%%%%%%%%%%%%%%%%%%%%%%%%%%%%%%%%%%%%%%%%%%%%%%%%%%%%%
%%%%%%%%%%%%%%%%%%%%%%%%%%%%%%%%%%%%%%%%%%%%%%%%%%%%%%%%%%%%

%%%%%%%%%%%%%%%%%% %%%%%%%%%%%%%%%%%%%%%%%% %%%%%%%%%%%% 

%\newpage
		%%%%%%%%%%%%%%%%%% TABLES %%%%%%%%%%%%%%%%%%%%%

%%%%%%%%%%%%%%%%%%  TABLE I %%%%%%%%%%%%%%%%%%%%%%%%
\begin{table}
\begin{tabular}{cccccc}
Group &\multicolumn{3}{c}{Embedding factors}& 
\multicolumn{2 }{c}{$\Delta\alpha$ } \\
      &\quad $c_1$\quad  & \quad $c_2$ \quad & $c_3$ & MSSM \qquad &
      \qquad  SM \qquad \\
\hline
SU(5); SO(10); E$_6$; [SU(3)]$^3$; etc. & $3\over 5$ & $1$ & $1$ &0.92784 &  41.1298 \\
SU(5)$\times$ SU(5); 
SO(10)$\times$ SO(10) & $3\over 13$  & $1$       & $1\over 2$ & 56.2556 & 122.675 \\ \
[SU(3)]$^4$ & $3\over 5$   & $1$       & $1\over 2$ & 68.5661 & 174.603 \\ \
[SU(4)]$^3$ & $3\over 11$  & $1$       & $1$        & 53.398  & 117.118 \\ \
[SU(6)]$^3$ & $3\over 14$  & $1\over 3$& $1$        &-34.4107 &-62.9585 \\ \
[SU(6)]$^4$ & $3\over 19$  & $1\over 3$& $1\over 2$ &-2.44396 &-1.76904
\end{tabular}
\caption{$\Delta\alpha$ for several unified models calculated by using
central values of the experimental data.}
\end{table}

%%%%%%%%%%%%%%%%%%%%%%
%%%%%%%%%%%%%%%%%%  TABLE II %%%%%%%%%%%%%%%%%%%%%%%%
\begin{table}
\begin{tabular}{ccc}
Fields & $(\tilde{b}_Y,\tilde{b}_2,\tilde{b}_3)_{MSSM}$ & 
$(\tilde{b}_Y,\tilde{b}_2,\tilde{b}_3)_{SM}$ \\
\hline
Gauge bosons (G) & $ (0,-4,-6)$ & $( 0 , -7, - {21\over 2})$ \\
Higgs Fields (H) & $ (2,2,0) $  & $ ( {1\over 6}, {1\over 6}, 0 )$ \\
Fermions     (F) & $ ({20\over 3}, 4,4)\eta $  
		 & $({20\over 9},{4\over 3},{4\over 3})\eta $ \\
\hline
All fields ($\eta=3$)& $ (22,10,6) $  & $({41\over 6},-{17\over 6},-{13\over 2})$ 
\end{tabular}
\caption{Unnormalized beta functions for the different standard field
representations.}
\end{table}

%%%%%%%%%%%%%%%%%%%%%%
%%%%%%%%%%%%%%%%%%  TABLE III %%%%%%%%%%%%%%%%%%%%%%%%
\begin{table}
\begin{tabular}{ccccc}
Model & High dimensional fields & $\Lambda$\ (GeV) & $\mu_0$\ (GeV) & $\alpha^{-1}$ \\
\hline
		%%%%%%
SU(5) Class &
	  G + H (+ F) & 1.395 $\times 10^{16}$ &
\begin{tabular}[c]{c} (9.239) \\[-1.5ex] (5.876)  \end{tabular} $\times 10^{15}$
& 24.571 - 0.387 $\eta$  \\
\hline
		%%%%%%
 [SU(5)]$^2$ Class 
 	 & All fields & 1.081 $\times 10^{11}$ &
\begin{tabular}[c]{c} 1.1(03) \\[-1.5ex] 1.1(13)  \end{tabular} $\times 10^{10}$
 & 1.873 \\
\hline
		%%%%%%

 [SU(3)]$^4$ 
     & All fields & 7.386 $\times 10^{12}$ &
\begin{tabular}[c]{c} 8.2(04) \\[-1.5ex] 8.2(78)\end{tabular} $\times 10^{11}$
 & 3.602   \\ 
\hline
		%%%%%%
 [SU(4)]$^3$  
	& All fields & 2.172 $\times 10^{10}$ &
\begin{tabular}[c]{c}  2.(389) \\[-1.5ex] 2.(409) \end{tabular} $\times 10^{9}$
 & 4.257     \\ 
\hline
		%%%%%%
 [SU(6)]$^3$   
 	& G + F & 7.758 $\times 10^{10}$ &
\begin{tabular}[c]{c} (6.985) \\[-1.5ex] (7.035) \end{tabular} $\times 10^{9}$
 & 1.211  \\ 
 \hline
		%%%%%%
[SU(6)]$^4$   

       & G + F & 1.866 $\times 10^{11}$ &
\begin{tabular}[c]{c} 2.4(47) \\[-1.5ex] 2.4(77) \end{tabular} $\times 10^{10}$
 & 3.959  	 
\end{tabular}
\caption{Predictions of the MSSM  for $\Lambda$, $\mu_0$ and $\alpha^{-1}$ for
several models. The upper (lower) value of $\mu_0$  comes  from the power law
(logarithmic)  running. The  brackets in $\mu_0$ values stand to stress the
numerical position where the   logarithmic running prediction starts to deviate
from the power law. The notation of high dimensional fields
is as in Table II. }
\end{table}

%%%%%%%%%%%%%%%%%%%%%%%%%%%%%%%%%%%%%%%%%%%%%%%%%%%%%%%%%%%%%%%%%%

%%%%%%%%%%%%%%%%%%%%%%%%%%%%%%%%%%%%%%%%%%%%%%%%%%%%%
%%%%%%%%%%%%%%%%%%%%%%
%%%%%%%%%%%%%%%%%%  TABLE IV %%%%%%%%%%%%%%%%%%%%%%%%
\begin{table}
\begin{tabular}{c c c }
Fields      & $(\tilde{b}'_1,\tilde{b}'_2,\tilde{b}'_3)$ & 
${5\over 4}({\bf u}\times{\bf b})\cdot\tilde{\bf b}'$ \\
\hline
Gluons       & $(0,0,-6)$        	& 42  \\
Left bosons  & $(0,-4,0)$		& -48 \\
Right Bosons & $(-{12\over 5},0,0)$ 	& 12  \\
$B-L$ Boson  & $(0,0, 0)$		& 0   \\ 
\hline
All bosons   & $(-{12\over 5}-4,-6)$	& 6   \\
\hline
$L(1,2,1,-1)$       & $({3\over 5},1,0)$	& 9    \\
$L^c(1,1,2,1)$       & $({6\over 5},0,0)$	& -6   \\
$Q(3,2,1,{1\over 3})$ & $({1\over 5},3,2)$	& 21   \\
$Q^c(3,1,2,-{1\over 3})$ & $(2,0,2)$		&-24   \\
\hline
All fermions &  $(4,4,4)\eta$ 			& 0    \\
\hline
$\Delta_R(1,1,3,-2)+ \bar{\Delta}_R(1,1,3,2)$ 
	          & $(12,0,0)$			& -60  \\
$\phi(1,2,2,0)$	     & $({6\over 5},2,0)$		& 18   \\
$\chi_L(1,2,1,1) + \bar\chi_L(1,2,1,-1)$ & $({6\over 5},2,0)$ & 18 \\
$\chi_R(1,1,2,1) + \bar\chi_R(1,1,2,-1)$ & $({12\over 5},0,0)$ & -12 \\
\hline
All Scalars  & 
$({12\over 5}(5 n_T + n_R) + {6\over 5}(n_B+ n_L), 2 (n_B+ n_L),0)$ 
		& $18 (n_B +n_L)  - 12(5 n_T + n_R)$
\end{tabular} 
\caption{Beta functions for the high dimensional
fields in the LRM. Their contribution to
the splitting of the unification scale has been normalized  in order to
simplify matters. The fermionic and scalar representations under the LRM are as
it is indicated.}
\end{table}
\newpage

%%%%%%%%%%%%%%%%%% %%%%%%%%%%%%%%%%%% %%%%%%%%%%%%%%%%%% 

%%%%%%%%%%%%%%%%%%%%%%% FGURES   %%%%%%%%%%%%%%%%%%%%%

 %%%%%%%%%%%%%%%%%%%%%
\begin{figure}%[ht]
\centerline{
\epsfxsize=250pt
\epsfbox{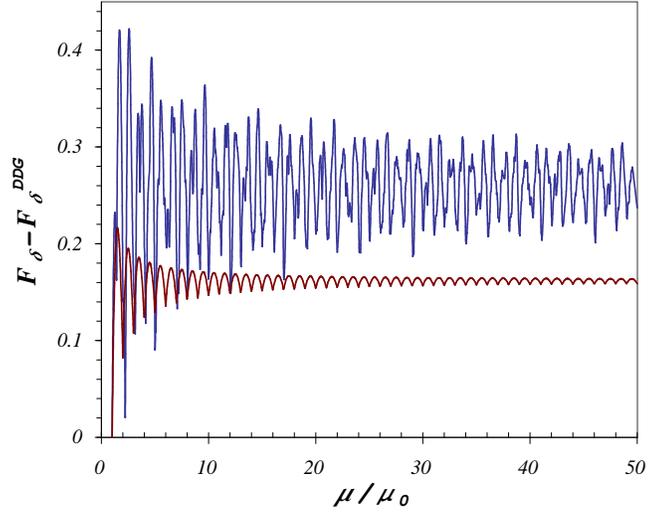}
}
\caption{Numerical difference among the logarithmic ($F_\delta$) 
and power law ($F_\delta^{DDG}$) expressions discussed in the main text 
plotted as a function of the ratio $\mu/\mu_0$. The upper (lower) 
curve corresponds to two (one) extra dimensions.}
\end{figure}
%%%%%%%%%%%%%%%%%%%%
\vskip2em

\begin{figure}%[ht]
\centerline{
\epsfxsize=250pt
\epsfbox{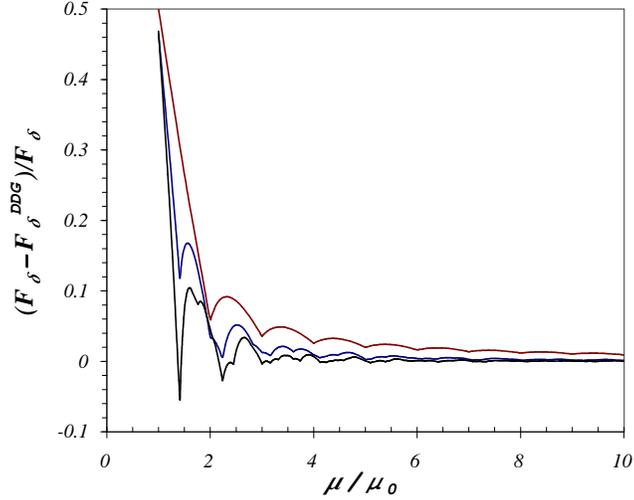}
}
\caption{Suppression of the 
difference $F_\delta - F_\delta^{DDG}$ by $F_\delta$  displayed in terms
of the  ratio $\mu/\mu_0$. Upper (middle) curve corresponds to $\delta=1$
($\delta=2$). The inferior curve are results for $\delta=3$. }
\end{figure}
%%%%%%%%%%%%%%%%%%%%
\vskip2em

\begin{figure}%[ht]
\centerline{
\epsfxsize=250pt
\epsfbox{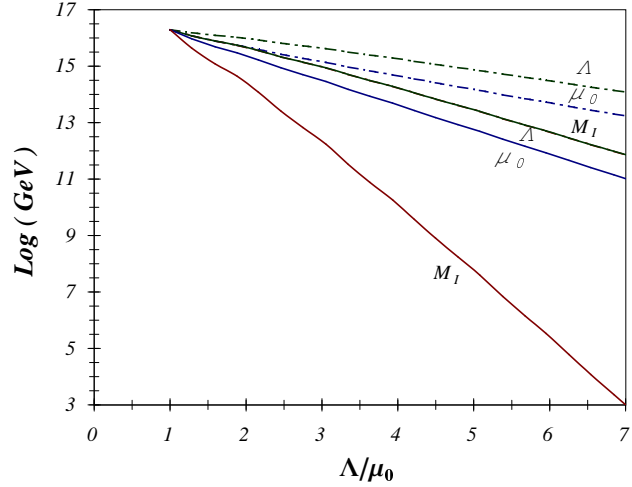}
}
\caption{The predictions for the unification, compactification and the
intermediate scales $\Lambda, \mu_0, M_I$ for the case where the
KK modes  of the gauge bosons of the LRM are present, are plotted against
of the ratio $\Lambda/\mu_0$.
Solid (dashed) lines correspond to the  case where $n_T=1$ and $n_B=2$
($n_B=1$) without scalar doublets.  The superposition of middle curves is
an
accidental effect. The extreme left corner point corresponds to the
standard MSSM unification.}
\end{figure}
%%%%%%%%%%%%%%%%%%%%
\vskip1em

%%%%%%%%%%%%%%%%%%%%%
\begin{figure}%[ht]
\centerline{
\epsfxsize=250pt
\epsfbox{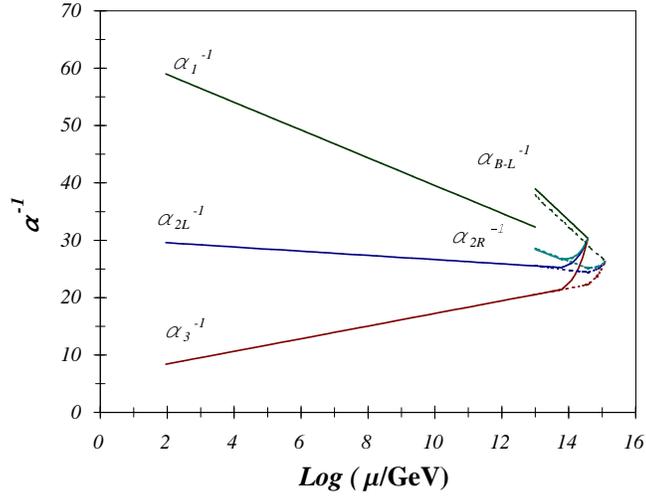}
}
 \caption{Running of the gauge couplings in the minimal  (solid
lines): $n_T=n_B=1$  and
next to minimal (dashed lines): $n_T=1$; $n_B=2$; left right models 
with
$M_{W_R}=10^{13}$ GeV;
assuming that only the gauge bosons develope KK modes.}
 \end{figure}
%%%%%%%%%%%%%%%%%%%%
\vskip1em

%%%%%%%%%%%%%%%%%%%%%
\begin{figure}%[ht]
\centerline{
\epsfxsize=250pt
\epsfbox{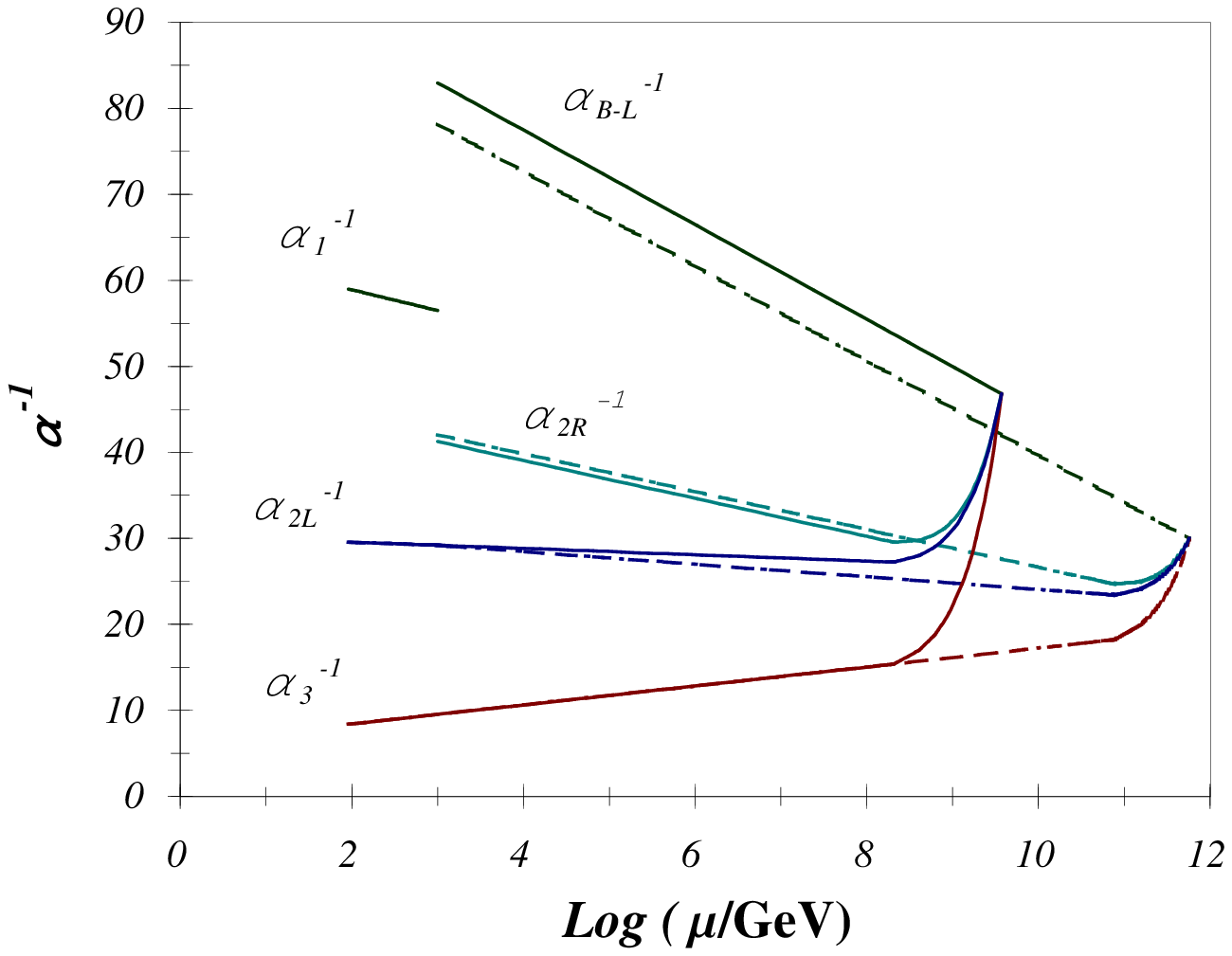}
}
 \caption{Running of the gauge couplings in the minimal  (solid
lines) and
next to minimal (dashed lines) left right models 
for $M_{W_R}=1$ TeV;
assuming that only the gauge bosons propagate into the bulk.} 
 \end{figure}
%%%%%%%%%%%%%%%%%%%% 
%\vskip1em


\begin{thebibliography}{99}
\bibitem{comment} 
This is a conventional normalization. Actually, $c_2^{-1}$  represents the
number of $SU(2)_L$ doublets in the fundamental representation of $G$. For
instance the actual values for SO(10) are  $(c_1,c_2,c_3) = ({3\over
10},{1\over 2},{1\over 2}) ={1\over 2}({3\over 5},1,1)$. The extra factor is
then absorved into the unified coupling $\alpha$.
\end{thebibliography}
\end{document}